\documentstyle[12pt]{article}

\begin{document}

%=======================================================================
%   BEGIN TITLE PAGE
%=======================================================================
\baselineskip=18pt % Title Page

\begin{titlepage}

\begin{flushright}
UR-1407, ER-40685-854 \\
March 1995
\end{flushright}

\medskip

\begin{center}

{\bf DISCOVERING THE HIGGS BOSONS \\
OF MINIMAL SUPERSYMMETRY WITH MUONS}

\vspace{0.24in}

CHUNG KAO\footnote{Internet Address: KAO@URHEP.PAS.ROCHESTER.EDU}

{\sl Department of Physics and Astronomy, University of Rochester \\
Rochester, NY 14627, USA}

\bigskip

NIKITA STEPANOV\footnote{Internet Address: STEPANOV@CERNVM.CERN.CH}

{\sl Institute of Theoretical and Experimental Physics \\
Moscow, Russia }

\end{center}

\vspace{0.24in}

\begin{abstract}

The prospects of detecting neutral Higgs bosons
in the minimal supersymmetric model
via their decays into muon pairs at the LHC are investigated.
The CMS detector performance is adopted for a realistic study of observability.
It is found that the muon pair decay mode
might provide a very promising channel
to search for the neutral Higgs bosons of minimal supersymmetry.
This channel will allow precise mass reconstruction
for neutral Higgs bosons.

\end{abstract}

\end{titlepage}

%==============================================================================
%   BEGIN TEXT
%==============================================================================
\baselineskip=24pt % Double Space

%------------------------------------------------------------------------------
% 1. Introduction
%------------------------------------------------------------------------------
\section{Introduction}

\medskip

One of the most important experimental goals
of future hadron supercolliders such as the CERN Large Hadron Collider (LHC)
is to unravel the mystery of electroweak symmetry breaking,
to produce and to detect the Higgs bosons or to prove their non-existence.
In the Standard Model (SM) of electroweak interactions,
only one Higgs doublet is required to generate masses for fermions
as well as gauge bosons. One neutral CP-even Higgs boson ($H^0$)
appears after spontaneous symmetry breaking.
Various extensions of the SM have more complicated Higgs sectors
and lead to additional physical spin zero fields \cite{GUIDE}.

The minimal supersymmetric extension of the Standard Model (MSSM) \cite{MSSM}
has two Higgs doublets with vacuum expectation values $v_1$ and $v_2$.
After spontaneous symmetry breaking, there remain five physical Higgs bosons:
a pair of singly charged Higgs bosons $H^{\pm}$,
two neutral CP-even scalars $H$ (heavier) and $h$ (lighter),
and a neutral CP-odd pseudoscalar $A$.
The Higgs sector is strongly constrained by supersymmetry
so that at the tree level, all Higgs boson masses and couplings are
determined by just two independent parameters,
which are commonly chosen to be the mass of the CP-odd pseudoscalar ($M_A$)
and $\tan \beta \equiv v_2/v_1$.

Since the top quark is very heavy \cite{CDF,D0},
radiative corrections from large $t$-quark Yukawa couplings substantially
modify the tree level formulae for masses and mixing patterns in the Higgs
sector.
Several groups have reevaluated prospects for the detection
of MSSM Higgs bosons at future hadron colliders.
Most studies have focused on various SM decay modes
for the neutral Higgs bosons,
$\phi \to \gamma\gamma$ and $\phi \to ZZ$ or $ZZ^*\to 4l$
($\phi = H, h$ and $A$);
as well as for the charged Higgs boson \cite{HC},
that are detectable above background.
Parameters were selected such that
supersymmetric particle (SUSY particle) masses were large
so that Higgs boson decays to SUSY particles
were kinematically forbidden \cite{HGG}-\cite{KZ}.
A region of parameter space roughly spanning pseudoscalar Higgs mass
$M_A \sim 100-300$ GeV and ratio of Higgs VEV's $\tan\beta \sim 4-10$
where none of the SM decay modes were detectable.
For large $\tan\beta$, the $\tau\bar{\tau}$ decay mode \cite{KZ,Unal,CMS}
has been suggested to be a promising discovery channel.
Recently, it has been argued that
neutral Higgs bosons might be observable via
their $b\bar{b}$ decays \cite{Dandi,DGV,SMW} in a large region
of the ($M_A,\tan\beta$) plane,
provided that sufficient $b$-tagging capability can be achieved.
Signals for invisible decays of Higgs bosons
have also been considered \cite{HDDT}-\cite{Jack}.
In addition, there are regions of parameter space where rates for Higgs boson
decays to SUSY particles are large and dominant.
While these decays reduce the rates for SM signatures, making
conventional detection of Higgs bosons even more difficult,
they also open up a number of new promising modes for Higgs detection
\cite{Z2Z2}.

In this paper, the prospects of discovering
the neutral Higgs bosons in the MSSM
via their decays into muon pairs\footnote{This discovery channel may not be
useful to search for the SM Higgs boson \cite{GUIDE}.}
at the LHC are investigated.
Parton level calculations are presented in Section II.
The CMS detector performance is adopted for a realistic study of observability.
Results from realistic simulations are discussed in Section 3
and promising conclusions are drawn in Section 4.

%------------------------------------------------------------------------------
% 2. Parton Level Calculations
%------------------------------------------------------------------------------
\section{Parton Level Calculations}

In our analysis, the cross section of $pp \to \phi \to \mu\bar{\mu} +X$
is evaluated from the Higgs boson cross section $\sigma(pp \to \phi +X)$
multiplied with the branching fraction of the Higgs decay into muon pairs
$B(\phi \to \mu\bar{\mu})$.
The parton distribution functions of CTEQ2L \cite{CTEQ}
are chosen to evaluate the cross section of $pp \to \phi +X$
with $\Lambda_4 = 0.190$ GeV and $Q^2 = M_\phi^2 +P_T^2$,
where $P_T$ is the transverse momentum of the Higgs bosons.
We take $M_Z = 91.187$ GeV, $\sin^2\theta_W = 0.2319$,
$M_W = M_Z \cos\theta_W$,
$m_b = 4.7$ GeV, and $m_t = 175$ GeV.
We have included one loop corrections from top and bottom
Yukawa interactions to the Higgs masses and couplings
using the effective potential \cite{OYY}-\cite{Haber}
with $A_t = A_b = 0$.
The contributions from the D-terms are usually small \cite{Z2Z2},
therefore, they are not included.
We consider two sets of parameters similar
to those discussed in Ref. \cite{Z2Z2},
(a) $m_{\tilde{g}} = m_{\tilde{q}} = \mu = 1000$ GeV, such that
the Higgs boson decays to SUSY particles are kinematically forbidden;
and (b) $m_{\tilde{g}} = m_{\tilde{q}} = \mu = 300$ GeV, such that
the Higgs boson decays to SUSY particles are large and dominant
when $\tan\beta$ is less than about 10.
The K-Factors are not included for the signal or the background.

%   Production Cross Section

At the LHC energy, the SM Higgs boson ($H^0$) is produced dominantly
from gluon fusion \cite{gluon},
and from vector boson fusion \cite{Cahn}-\cite{Kane} if the Higgs boson
is heavy.
In the MSSM, gluon fusion ($gg \to \phi$) is the major source
of neutral Higgs bosons for $\tan\beta$ less than about 4.
If $\tan\beta$ is larger than about 10,
neutral Higgs bosons in the MSSM are dominantly produced
from $b$-quark fusion ($b\bar{b} \to \phi$) \cite{Duane}.
We have evaluated the cross section of Higgs bosons in $pp$ collisions
$\sigma(pp \to \phi +X)$,  with two dominant subprocesses:
$gg \to \phi$ and $gg \to \phi b\bar{b}$.
The cross section of $gg \to \phi b\bar{b}$
is a good approximation to the `exact' cross section \cite{Duane}
of $b\bar{b} \to \phi$ for $M_\phi$ less than about 500 GeV.
In addition, the subprocesses $gg \to \phi b\bar{b}$ and $gg \to g\phi$
are complementary to each other for producing large $P_T$ Higgs bosons
at future hadron colliders \cite{EHSV,Uli,Kao}.
Since the Yukawa couplings of $\phi b\bar{b}$ are enhanced by $1/\cos\beta$,
the production rate of neutral Higgs bosons is usually
enhanced with large $\tan\beta$.
For $M_A$ larger than about 150 GeV, the couplings of the lighter scalar $h$
to gauge bosons and fermions become close to those of the SM Higgs boson,
therefore, gluon fusion is the major source of the $h$
even if $\tan\beta$ is large.

%   Decay Branching Fraction

If the $b\bar{b}$ mode dominates Higgs decays,
the branching fraction of $\phi \to \mu\bar{\mu}$ is about $(m_\mu/m_b)^2$.
It has been found that QCD radiative corrections reduce
the decay width of $\phi \to b\bar{b}$ by a factor of about 2
\cite{Braaten,Manuel}.
Therefore, the branching fraction of $\phi \to \mu\bar{\mu}$
is about $2 \times 10^{-4}$ when the $b\bar{b}$ mode dominates.
Let's consider a parameter space of ($M_A,\tan\beta$) with
50 GeV $\le M_A \le$ 500 GeV and $1 \le \tan\beta \le 35$.
The branching fraction of $B(h \to \mu\bar{\mu})$ in the whole parameter space
as well as $B(A \to \mu\bar{\mu})$ and $B(H \to \mu\bar{\mu})$
with $\tan\beta \matrix{>\cr\noalign{\vskip-7pt}\sim\cr} 10$
is always about $2 \times 10^{-4}$,
even when $A$ and $H$ can decay into SUSY particles.
For $M_A$ less than about 80 GeV, the $H$ decays dominantly into
$hh$, $AA$ and $ZA$.

Fig. 1 shows the cross section of $pp \to \phi \to \mu\bar{\mu} +X$
as a function of $M_A$ for various values of $\tan\beta$.
We have taken $m_{\tilde{q}} = \mu = 1$ TeV.
For $\tan\beta \matrix{>\cr\noalign{\vskip-7pt}\sim\cr} 10$,
the production cross section is always enhanced.
As expected, the cross section of $pp \to h \to \mu\bar{\mu} +X$
does not change much with $\tan\beta$ for $M_A > 150$ GeV.
Also shown is the same cross section for the SM Higgs boson $H^0$
with $M_{H^0} = M_A$.
For $M_{H^0} > 140$ GeV, the SM $H^0$ mainly decays into gauge bosons,
therefore, the branching fraction $B(H^0 \to \mu\bar{\mu})$ drops sharply.

We define the signal to be observable if the $99\%$
confidence level upper limit on the background is smaller than the
corresponding lower limit on the signal plus background \cite{HGG,Brown},
namely,
\begin{eqnarray}
L(\sigma_s+\sigma_b)-N\sqrt{L(\sigma_s+\sigma_b)} & > &
L\sigma_b+N\sqrt{L\sigma_b} \nonumber \\
\sigma_s & > & \frac{N^2}{L}[1+2\sqrt{L\sigma_b}/N]
\end{eqnarray}
where $L$ is the integrated luminosity, and $\sigma_b$ is the background cross
section within a bin of width $\pm\Delta M_{\mu\bar{\mu}}$ centered
at $M_\phi$; $N = 2.32$ corresponds to a $99\%$ confidence level
and $N = 2.5$  corresponds to a 5$\sigma$ signal.

To study the observability of the muon discovery mode,
we consider the background from the Drell-Yan (DY) process,
$q\bar{q} \to Z,\gamma \to \mu\bar{\mu}$, which is the dominant background.
We take $\Delta M_{\mu\bar{\mu}}$ to be the larger of
the CMS muon mass resolution or the Higgs boson width.
The minimal cuts applied are (1) $p_T(\mu) > 10$ GeV and
(2) $|\eta(\mu)| < 2.5$, for both the signal and background.
More details about the background will be discussed in next section.

The 5$\sigma$ discovery contours at $\sqrt{s} =$ 14 TeV
are shown in Fig. 2, for
(a) $L = 100$ fb$^{-1}$, $\mu = m_{\tilde{q}} = 1000$ GeV,
(b) $L = 100$ fb$^{-1}$, $\mu = m_{\tilde{q}} =  300$ GeV, and
(c) $L =  10$ fb$^{-1}$, $\mu = m_{\tilde{q}} = 1000$ GeV.
The $A$ might be detectable in a large region of parameter space
with $\tan\beta$ away from one.
The $H$ might be observable in a region with $M_A >$ 120 GeV
and $\tan\beta \matrix{>\cr\noalign{\vskip-7pt}\sim\cr} 10$.
The $h$ might be observable in a region with $M_A < 120$ GeV
and $\tan\beta \matrix{>\cr\noalign{\vskip-7pt}\sim\cr} 4$.
This channel is affected by the SUSY decay modes only slightly for $H$ and
$h$ \cite{Mike}.
The lighter top squarks will make the $H$ and $h$ lighter and enhance
the $Hb\bar{b}$ coupling while reducing the $hb\bar{b}$ coupling.
Therefore, the discovery region of $H \to \mu\bar{\mu}$ is slightly enlarged
for a smaller $M_A$,
but the observable region of $h \to \mu\bar{\mu}$ is slightly reduced.

%------------------------------------------------------------------------------
% 3. Signal and backgrounds estimates
%------------------------------------------------------------------------------
\section{Realistic Simulations}

\medskip

Like the $\gamma \gamma$ mode, the discovery potential of the
$\mu \mu$ channel is very sensitive to the detector performance.
Therefore, one needs to fix some detector model for quantitative estimates.
The CMS detector performance parameters
are used in our analysis to estimate the signal and backgrounds.
The CMS will be one of two general purpose LHC detectors, and
the very precise muon momentum reconstruction is one of its main
parameters, which define the detector design \cite{CMS}.
Thus, it will be an excellent detector to look for the narrow di-muon
resonances.

\subsection{Calculation tools}

We use PYTHIA 5.7 and  JETSET 7.4  generators \cite{PYTHIA}
with the CTEQ2L \cite{CTEQ} parton distribution functions
to simulate events on the particle level.
In the case of DY background near the $Z$-peak, QED corrections
are very important and have been taken into account
by treating the $Z \to \mu\bar{\mu}$ decay
with the computer program PHOTOS \cite{PHOTOS}.

In addition to $gg \to \phi$, the new debugged PYTHIA \cite{PYT1}
allows one to use $gg \to \phi b\bar{b}$ process
for the production of all neutral MSSM Higgs bosons.
The generated kinematics and
total cross sections are in reasonable agreement
with analytical parton level calculations.
For our purpose we adapted in PYTHIA
the program calculating 1-loop corrected masses and couplings \cite{Fabio}.
The PYTHIA/JETSET/PHOTOS outputs are processed
with the CMSJET program \cite{CMSJET}.
It is developed for fast simulations of "realistic" CMS detector response.
The resolution effects are taken into account
by using the parametrizations obtained
from the detailed GEANT \cite{GEANT} simulations.
CMSJET includes also some analysis programs, in particular,
a set of jet reconstruction algorithms, etc.

For future analysis,
all muons with transverse momentum $p_{T}(\mu) \geq 5$ GeV
and pseudorapidity $|\eta(\mu)| \leq 2.4$;
missing transverse momentum $\ \hbox{{$p$}\kern-.43em\hbox{/}}_T$;
central jets with transverse energy $E_{T} \geq 40$ GeV
and $|\eta| \leq 2.4$ are reconstructed and stored for each generated event.

\subsection{Backgrounds and optimal kinematical cuts}

There are several SM physical processes
that generate sizeable muon pairs\footnote{SUSY backgrounds
from the lightest chargino pairs and
from the second lightest neutralino decays are currently under investigation.
In our analysis, we assume that they could be reduced significantly
with a cut on missing transverse energy, similar to the $t\bar{t}$ decays.}.
However, analyzing them process by process, one can conclude,
that there is only one dominant subprocess
in the interesting mass range
(50 GeV $\leq M_{\mu\bar{\mu}} \leq 500$ GeV),
which is the DY muon pair production \cite{Nikita}.
For low
$M_{\mu\bar{\mu}} \matrix{<\cr\noalign{\vskip-7pt}\sim\cr} 100$ GeV
the $b\bar{b}$ background is also potentially dangerous,
but it can be reduced easily well below the
DY process with isolation cuts on muons.
The background from top quark pairs ($t\bar{t}$) is negligible in the region
$| M_{\mu\bar{\mu}} - M_{Z} | \leq$ 30 GeV.
It is about 20 times smaller than the DY
for $M_{\mu\bar{\mu}} \matrix{<\cr\noalign{\vskip-7pt}\sim\cr} 150$ GeV
and begins to compete with the DY for the highest $M_{\mu\bar{\mu}}$.
About 90$\%$ of muon pairs in $t\bar{t}$ production are generated
from the decay chain
$t \bar{t} \to W W b\bar{b} \to \mu\bar{\nu}\bar{\mu}\nu b\bar{b}$.
For comparison, the event rate from $WW$ is 3-5 times lower than that
from $t \bar{t}$.

To optimize the kinematical cuts, sizeable ($\sim 10^{6}$) background
event samples are generated using the production chain described above.
Our analysis indicates, that, in fact,
after $M_{\mu\bar{\mu}}$ is constrained
in a small bin around some fixed $M_\phi$
(we are looking for the narrow states),
there are no further suitable kinematical cuts to suppress
the dominant DY background.
So here some minimal cuts, providing good
signal efficiency and acceptance are used: $p_{T}(\mu) \geq 10$ GeV
with $\eta(\mu) \leq 2.4$.
The $t\bar{t}$ background can be reduced nearly by a factor of 5
with additional cuts:
missing transverse energy $ \ \hbox{{$E$}\kern-0.60em\hbox{/}}_T \leq 50$ GeV,
$N_{jet} \leq 1$ for jets with $E_{T} \geq 40$ GeV,
and $|\eta(jet)| \leq 2.4$.
The muon isolation reduces the $b\bar{b}$ background by
a factor of about 100.

\subsection{Results}

The invariant mass distribution ($d\sigma/dM_{\mu\bar{\mu}}$)
of the backgrounds from (a) Drell-Yan, (b) $W^+W^-$, (c) $t\bar{t}$,
and (d) $b\bar{b}$, are shown in Fig. 3.

Fig. 4 shows the $5 \sigma$ significance contour for CMS detector and
$L = 100$ fb$^{-1}$ with minimal cuts discussed above and isolation
criteria applied for low masses ($M_{\mu\bar{mu}} \leq 85$ GeV).
In this figure, we have considered the total cross section
of $pp \to \phi \to \mu\mu +X$, summed over the $h$, the $H$ and the $A$.
For $M_{A} \matrix{<\cr\noalign{\vskip-7pt}\sim\cr} 200$ GeV
and $\tan\beta \geq 25 - 30$,
an integrated luminosity of 10 fb$^{-1}$ would be sufficient
to obtain Higgs signals
with a statistical significance larger than 7.
But in the region close to the $Z$- peak, the signal is marginal
despite the large significance, because it appears on the shoulder
of the huge $Z$- peak. Only an adequate subtraction procedure,
if it would be possible, enables one to extract the signal in this region.

Despite the high di-muon mass resolution of CMS detector
(about 0.5\% for $M_{\mu\bar{\mu}} \simeq 100$ GeV and
better than 2\%
for $M_{\mu\bar{\mu}} \matrix{<\cr\noalign{\vskip-7pt}\sim\cr} 500$ GeV),
for large $\tan\beta \sim 30$ one can obtain the unresolved (h,A) signal
for $M_{A} \matrix{<\cr\noalign{\vskip-7pt}\sim\cr} 120$ GeV
or (H,A) signal for higher $M_{A}$,
because the mass difference between the (h,A) or (H,A) Higgs bosons is
several hundreds MeV, i.e. a factor of 10 less than Higgs widths.
Only for the lowest accessible $\tan\beta \sim 10-15$ it seems to be possible
to resolve the signals for two resonances.
But the signal become not so significant.
Fig. 5 illustrates the signal over the backgrounds
as function of $M_{\mu\bar{\mu}}$
at $\sqrt{s} = 14$ TeV with an integrated luminosity of 100 fb$^{-1}$,
$M_A = 80,150$ GeV, and $\tan\beta = 30$.
This figure is generated from a simulation with the CMS performance
and $m_{\tilde{q}} = \mu =$ 1000 GeV.

%-----------------------------------------------------------------------
%   Conclusions
%-----------------------------------------------------------------------
\bigskip
\section{Conclusions}

It is found that the muon pair decay mode
can be a very promising channel
to discover the neutral Higgs bosons of minimal supersymmetry.
The discovery region of the $\mu\bar{\mu}$ might be slightly smaller
than the $\tau\tau$ channel but it will allow precise reconstruction
for the Higgs boson masses.
The $A$ might be observable in a large region of parameter space
with $\tan\beta$ away from one.
The $H$ might be detectable in a large region with $M_A >$ 120 GeV
and $\tan\beta \matrix{>\cr\noalign{\vskip-7pt}\sim\cr} 10$.
The $h$ might be observable in a region with $M_A < 120$ GeV
and $\tan\beta \matrix{>\cr\noalign{\vskip-7pt}\sim\cr} 4$.

For $M_A \leq 200$ GeV and $\tan\beta > 25$, $L =$ 10 fb$^{-1}$
would be enough to obtain Higgs boson signals
with a statistical significance larger than 7.
For $M_{\mu\bar{\mu}}$ close to the $M_Z$, the signal is marginal
despite the large significance, because it appears on the shoulder
of the huge $Z$ peak. Adequate subtraction procedures
are required to extract the signal in this region.
One attractive possibility is to tag the $b$-jets accompanying the
$\mu\bar{\mu}$ from the Higgs decay,
since the production rate of Higgs bosons via $gg \to \phi b\bar{b}$ is large.
With high $b$-tagging efficiency and purity,
we can reduce the dominant DY background
to the level of $Z + b\bar{b}$ production.
In this case, the $t\bar{t}$ background may become dangerous,
and the additional cuts discussed above may be crucial.
This signature is currently under study.

%-----------------------------------------------------------------------
%   THE ACKNOWLEDGEMENTS
%-----------------------------------------------------------------------
\bigskip

\noindent{\large\bf Acknowledgements}

\medskip

We would like to thank Howie Baer and Xerxes Tata
for beneficial discussions and comments, and especially,
Duane Dicus for continuing encouragement as well as instructions.
This research was supported in part
by the US Department of Energy grant DE-FG02-91ER40685.

%-----------------------------------------------------------------------
%   THE BIBLIOGRAPHY
%-----------------------------------------------------------------------
\newpage
%

%-----------------------------------------------------------------------
%   FIGURE CAPTIONS
%-----------------------------------------------------------------------
\newpage
\noindent{\large\bf Figures}

\bigskip

FIG. 1
The total cross section of $pp \to \phi \to \mu\bar{\mu} +X$ in fb,
as a function of $M_A$, with $\sqrt{s} = 14$ TeV, $m_t = 175$ GeV,
$m_{\tilde{q}} = \mu =$ 1000 GeV, and $\tan \beta$ = 1, 3, 10, and 30,
for (a) the $H$, (b) the $h$ and (c) the $A$.
The cross section of $pp \to H^0 \to \mu\bar{\mu} +X$
for the SM Higgs boson is also presented with $M_{H^0} = M_A$.

\medskip

FIG. 2
The LHC discovery contours in the $M_A$ versus $\tan\beta$ plane,
for the $H$ (solid), the $h$ (dash) and the $A$ (dot).
Three cases are considered:
(a) $L = 100$ fb$^{-1}$, $\mu = m_{\tilde{q}} = 1000$ GeV,
(b) $L = 100$ fb$^{-1}$, $\mu = m_{\tilde{q}} =  300$ GeV, and
(c) $L =  10$ fb$^{-1}$, $\mu = m_{\tilde{q}} = 1000$ GeV.
Other parameters are the same as in Fig. 1.
The discovery regions are on the upper plane of the the parameter space.

\medskip

FIG. 3 Invariant mass distribution ($d\sigma/dM_{\mu\bar{\mu}}$)
of the background from (a) Drell-Yan, (b) $W^+W^-$, (c) $t\bar{t}$,
and (d)$b\bar{b}$.

\medskip

FIG. 4
The discovery contour in the $M_A$ versus $\tan\beta$ plane,
for the MSSM Higgs bosons via their muon pair decay mode,
at $\sqrt{s} = 14$ TeV with an integrated luminosity of 100 fb$^{-1}$.
This figure is generated from a simulation with the CMS performance
and $m_{\tilde{q}} = \mu =$ 1000 GeV.
We have considered the total cross section of $pp \to \phi \to \mu\mu +X$,
summed over the $H$, the $h$, and the $A$.
The QED radiative corrections to background from the Drell-Yan process
are included.
The discovery region is on the upper plane of the the parameter space.

\medskip

FIG. 5
Histograms of number of events as a function of $M_{\mu\bar{\mu}}$
for the signal and the background at $\sqrt{s} = 14$ TeV
with $L = 100$ fb$^{-1}$,
$M_A = 80,150$ GeV, and $\tan\beta = 30$.
This figure is generated from a simulation with the CMS performance
and $m_{\tilde{q}} = \mu =$ 1000 GeV.

%-----------------------------------------------------------------------
%   END DOCUMENT
%-----------------------------------------------------------------------
\end{document}